\documentclass[prc,twocolumn,showpacs,aps,superscriptaddress,floatfix]{revtex4}
\usepackage{graphicx}

\begin{document}

\title{Effect of phase relaxation on quantum superpositions in complex
collisions}

\author{L.~Benet}
\affiliation{Centro de Ciencias F\'{\i}sicas, National University of 
Mexico (UNAM), 62210--Cuernavaca, Mor., Mexico}
\author{S.Yu.~Kun}
\affiliation{Centro de Ciencias F\'{\i}sicas, National University of 
Mexico (UNAM), 62210--Cuernavaca, Mor., Mexico}
\affiliation{Nonlinear Physics Center, RSPhysSE, The Australian National
University, Canberra ACT 0200, Australia}
\affiliation{Department of Theoretical Physics, RSPhysSE, The 
Australian National University, Canberra ACT 0200, Australia}
\author{Wang~Qi}
\affiliation{Institute of Modern Physics, Chinese Academy of Sciences, Lanzhou
730000, China}

\date{\today}

\begin{abstract}
  We study the effect of phase relaxation on coherent superpositions
  of rotating clockwise and anticlockwise wave packets in the regime
  of strongly overlapping resonances of the intermediate complex. Such
  highly excited deformed complexes may be created in binary
  collisions of heavy ions, molecules and atomic clusters. It is shown
  that phase relaxation leads to a reduction of the interference
  fringes, thus mimicking the effect of decoherence. This reduction is
  crucial for the determination of the phase--relaxation width from
  the data on the excitation function oscillations in heavy--ion
  collisions and bimolecular chemical reactions. The difference
  between the effects of phase relaxation and decoherence is
  discussed.
\end{abstract}


\maketitle

\section{Introduction}

For highly excited strongly interacting many--body systems the
independent particle picture has very limited
validity~\cite{Wigner72}. For high excitations, the interaction
results in a quick decay of
single--particle~\cite{BohWei88,AGKL97,GMGW98} as well as collective
modes~\cite{BohWei88,GMGW98}, which are not eigenstates of the
Hamiltonian of the system. This decay leads to the formation of
complicated many--body configurations. Each of these many--body states
is characterized by a uniform occupation of the accessible phase space
and sharing the energy between many particles of the system. The
characteristic time for the formation of such spatially extended
ergodic many--body states, independent of the initial conditions, is
given by the inverse spreading width, $\tau_{\rm erg}=
\hbar/\Gamma_{\rm spr}$~\cite{GMGW98}. Introduced by
Wigner~\cite{Wigner57}, $\Gamma_{\rm spr}$ also characterizes the
width of the distribution of the expansion coefficients of the
many--body eigenstates over a noninteracting mean--field basis. For
sufficiently high excitation energy the decay of single--particle
modes creates a linear superposition of a very large number of
many--body configurations. The question is whether the phase relations
between these individually ergodic many--body states in the
superposition may still preserve a memory of the way the system was
excited. This question is of fundamental importance for the study of
relaxation phenomena in nuclear, atomic, molecular and mesoscopic
many--body systems, and for many--qubit quantum computation. In
particular, if phase relaxation is longer than energy relaxation
$\hbar/\Gamma_{\rm spr}$, this can extend the time for quantum
computing~\cite{FKS05,marcSIGMA} beyond the quantum--chaos
border~\cite{GeoShep00,Shep01}.

To answer this question from first principles is difficult due to
computational limitations. Indeed, in order to solve the full quantum
many--body problem one would require a many--qubit quantum computer.
Therefore, the only currently available resort to search for possible
manifestations of long phase relaxation is the experiment, and a
careful data analysis. Nuclear systems are an ideal laboratory to
study many--body systems, since nuclear interactions are so strong
that external perturbations can often be neglected. In particular, the
analysis of Refs.~\cite{FKS05,marcSIGMA,marcPRC} of the data on
asymmetry around $90^\circ$ c.m. in angular distributions of
evaporated protons from heavy nuclei in nucleon--induced and
photonuclear reactions clearly indicates that phase relaxation time in
highly excited strongly interacting many--body systems can be up to
eight orders of magnitude~\cite{marcSIGMA} longer than energy
relaxation time $\hbar/\Gamma_{\rm spr}$. This reveals a new form of
matter--thermalized non--equilibrated matter introduced by one of us
in Refs.~\cite{Kun93,Kun94,Kun97}.

A more subtle indication of a slow phase relaxation is found in light
heavy--ion scattering~\cite{KVV98,KRV99,KVG01,KVCG02}. It manifests
itself in the formation of rotating wave packets whose spreading time,
which is given by the inverse phase--relaxation width, is much longer
than the energy relaxation time $\hbar/\Gamma_{\rm spr}$. The
approach~\cite{KRV99,KVG01,KVCG02} treats time--delayed light
heavy--ion scattering in terms of the formation and decay of
quasimolecular resonances~\cite{GPS95}. The highly excited coherently
rotating intermediate system has the energy of the intrinsic
excitation $\geq 10$~MeV. The intermediate system is in the regime of
strongly overlapping resonances. Therefore, this coherent rotation
does not originate from the discreteness of the spectrum, which is not
resolved, but is due to the spin off--diagonal correlations between
partial width amplitudes. Indeed, the period of the coherent rotation
is much shorter than the inverse level spacing of the intermediate
system. This reveals a new root, as compared to Bohr's correspondence
principle, for a quantum--classical transition in highly excited
many--body systems.

The width of the rotating wave packets is about $1/d+\beta
t/\hbar\approx 1/d+\beta\theta/\hbar\omega$, where $d$ is the
effective number of partial waves, $\beta$ is the spin off--diagonal
phase--relaxation width, $\omega$ is the angular velocity, and $t$ and
$\theta$ are the time and angle of the rotation, respectively. This
seemingly allows us to determine the wave--packet spreading rate from
the time power spectra at different scattering angles, which can be
reconstructed, for binary collisions, from the data on energy
fluctuations of the cross sections~\cite{KVG01,KVCG02}. These
fluctuations originate from the energy--fluctuating collision
amplitude corresponding to a resonance time--delayed process. However,
for elastic heavy--ion scattering, such energy fluctuations can be
typically observed only at backward angles $140^\circ\leq\theta\leq
180^\circ$, since for smaller $\theta$ the direct reaction
contribution grows exponentially and becomes much greater than the
time delayed one. Therefore, since at the initial moment of time the
deformed intermediate system is oriented in the forward direction
$\theta\approx 0^\circ$, the spreading of the wave packets within the
backward angular interval $140^\circ\leq\theta\leq 180^\circ$ can
hardly be detected reliably.  Moreover, the width of the wave packets
at this backward--angle range does not allow to determine $\beta$,
since the angular dispersion $1/d$ at $t=0$ is also unknown. In this
paper we show that both $d$ and $\beta$ can indeed be determined
unambiguously. This results from the strong sensitivity of the
interference fringes between the rotating clockwise and anticlockwise
wave packets for the backward angle range to the the phase relaxation
width: As $\beta$ grows, the interference fringes are suppressed more
strongly. The effect is considered in relation to heavy--ion
scattering and bimolecular chemical reactions. The difference between
the effects of phase relaxation and decoherence is discussed.

\section{The time power spectrum of the time--delayed collisions}

Following Ref.~\cite{KBCGH03}, we consider spinless collision partners
in the entrance and exit channels. The time and angle dependent
intensity of the decay (the time power spectrum), $P(t,\theta)$, is
given by the modulus square of the Fourier component of the
energy--fluctuating collision amplitude. It can be also expressed as
the Fourier component of the amplitude energy autocorrelation
function. $P(t,\theta )$ has been obtained in~\cite{KBCGH03} by
summing over a very large number of strongly overlapping resonance
levels, $\Gamma /D\gg 1$, where $\Gamma$ is the total decay width and
$D$ is the average level spacing of the intermediate complex. As a
result, after changing from summation over the resonance levels to
integration, which is a good approximation for $t\ll \hbar/D $,
$P(t,\theta )$ takes the form:
\begin{widetext}
\begin{equation}
 P(t,\theta ) \propto H(t)\exp(-\Gamma t/\hbar) 
  \sum_{JJ^\prime } (2J+1)(2J^\prime +1) [W(J) W(J^\prime )]^{1/2} 
  \exp[i(\Phi -\omega t) (J-J^\prime ) - \beta |J-J^\prime |t/\hbar ]
      P_J(\theta )P_{J^\prime}(\theta ).
\label{eq1}
\end{equation}
\end{widetext}
Here, $H(t)$ is the Heaviside step function, $\beta$ is the spin
phase--relaxation width, $\omega$ is the angular velocity of the
coherent rotation, $\Phi$ is the deflection given by the total spin
$J$ (in $\hbar$ units) derivative of the potential phase shifts
(direct reaction), and $P_J(\theta)$ are the Legendre polynomials. The
physical meaning of the inverse spin phase--relaxation width,
$\hbar/\beta$, is the characteristic time for the angular spreading of
the clockwise and anticlockwise rotating wave packets. The partial
average reaction probability is taken in the $J$--window form, $W(J)=
\langle |\delta S^J(E)|^2 \rangle \propto \exp[-(J-{\bar J})^2/d^2]$,
where ${\bar J}$ is the average spin and $d$ is the $J$--window width.

It should be noted that a similar effect of coherent rotation of the
nuclear molecule, as described by Eq.~(\ref{eq1}) with $\beta=0$, was
found in Ref.~\cite{Heinz84} for a scattering of heavy nuclei.
However, while Eq.~(\ref{eq1}) has been obtained for a large number of open
channels and strongly overlapping resonances of the intermediate
system~\cite{KBCGH03}, the time power spectrum in Ref.~\cite{Heinz84}
was derived from one open channel and one resonance pole form of the
$S$-matrix elements. Therefore the latter approach is not applicable
for the forthcoming analysis of the $^{12}$C+$^{24}$Mg system since,
for this system, the number of open channels is much greater than
one. Furthermore, the one--resonance pole form of the $S$-matrix
elements employed in Ref.~\cite{Heinz84} results in isolated
resonances with the energy spacing between them to be about
$\hbar\omega$. This is not the case for the $^{12}$C+$^{24}$Mg elastic
scattering analyzed in this paper. Indeed, the data on the excitation
function for the $^{12}$C+$^{24}$Mg elastic scattering at
$\theta=180^\circ$ reveal about 15 local maxima on the energy interval
$\Delta E_{cm}=13-22$~MeV~\cite{GhoshS87}. Then the
interpretation~\cite{Heinz84} would mean that average level spacing,
$\simeq 0.6$~MeV in our case, between the local maxima is given by
$\hbar\omega$. This, in turn, would reflect the quasiperiodic
behaviour of the cross section energy autocorrelation functions with
the period of about 0.6~MeV. This is inconsistent with the
data~\cite{GhoshS87} showing the quasiperiodicity of the cross section
with the period of about 2.9~MeV yielding $\hbar\omega=1.35-1.45$
MeV~\cite{KVG01,KVCG02}. Moreover, the interpretation~\cite{Heinz84}
implies that these local maxima in the excitation function should be
approximately equidistant. Instead, the data~\cite{Mermaz81,GhoshS87}
show big fluctuations, from 0.3~MeV to 1.3, for the energy spacing
between the nearest-neighboring local maxima in the excitation
function. Note that the indication against the interpretation in terms
of isolated resonances comes from the statistically insignificant
channel--channel correlations for the $^{12}$C+$^{24}$Mg elastic and
inelastic scattering at $\theta=180^\circ$~\cite{KVG01}. Finally, the
effect of coherent rotation~\cite{Heinz84} does not survive a
generalization~\cite{GreinAnnPhys} by taking into account intermediate
structure by means of inclusion of many isolated resonances in the
$S$--matrix elements. On the basis of the above arguments we believe
that the interpretation of the fine structure in the excitation
functions for the $^{12}$C+$^{24}$Mg scattering at $\theta=180^\circ$
in terms of overlapping, rather than isolated, resonances of the
intermediate system is a justified approach.

First, we calculate $P(t,\theta )$ for the set of parameters obtained
from the description~\cite{KVCG02} of the experimental cross section
energy autocorrelation functions~\cite{GhoshS87} for
$^{12}$C+$^{24}$Mg elastic and inelastic scattering at
$\theta=180^\circ$~\cite{Mermaz81}. For these collisions the analysis
of the oscillations in the cross section energy autocorrelation
functions~\cite{KVG01,KVCG02} indicates the formation of stable
rotational wave packets, in spite of the strong overlap of resonance
levels in the highly excited intermediate molecule. The set of
parameters is~\cite{KVCG02}: $\Phi=0$, $d=3$, ${\bar J}=14$,
$\beta=0.01$~MeV, $\hbar\omega =$1.45~MeV, and $\Gamma=0.3$~MeV.

\begin{figure}
\includegraphics[angle=270,width=8.5cm]{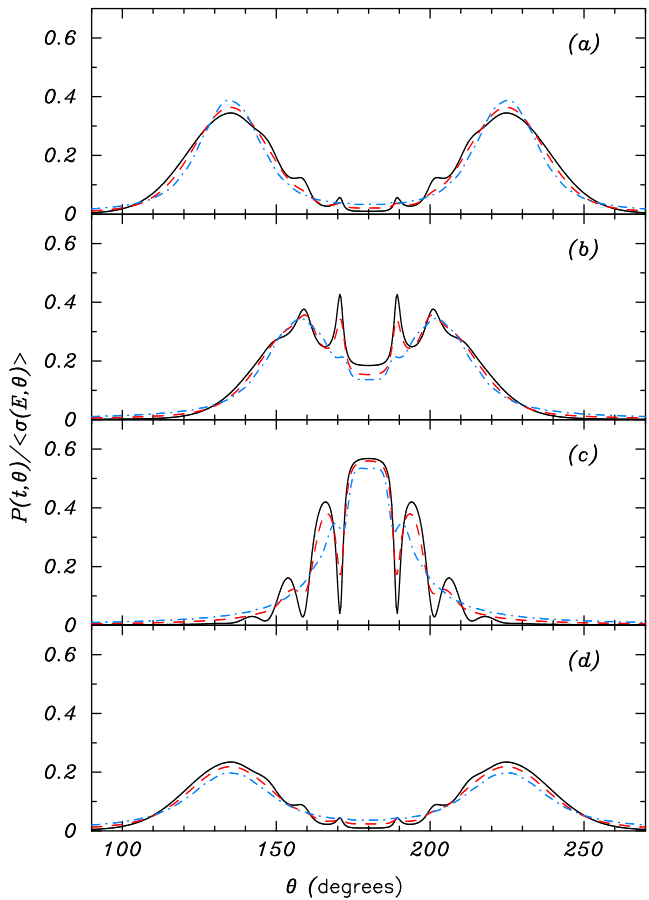}
\caption{\label{fig1} (Color online) Angular dependence of the decay
  intensity of the highly excited intermediate complex at different
  moments of time for different combinations of $\beta$ and $d$. Panel
  (a)~corresponds to time $t=3T/8$; (b)~$t=7T/16$; (c)~$t=T/2$;
  (d)~$t=5T/8$. Here, $T$ is the period of one complete revolution of
  the complex. Solid lines (black) correspond to $\beta=0.01$~MeV and
  $d=3$; dashed (red) to $\beta=0.04$~MeV and $d=4$; dashed--dotted
  (blue) to $\beta=0.075$~MeV and $d=8$. The bigger $\beta$ the
  smaller the interference fringe contrast is.}
\end{figure}

In Figs.~\ref{fig1} we plot the quantity $AP(t,\theta )/ \langle
\sigma(E,\theta) \rangle$ at four moments of time, with $T=2\pi/\omega$ being
the period of one complete revolution. Here, $\langle \sigma(E,\theta) \rangle
\propto \int_0^\infty dt \, P(t,\theta )$ is the energy--averaged differential
cross section for the time--delayed collision. In Figs.~\ref{fig1},
$P(t,\theta)$ is scaled with $\langle \sigma(E,\theta) \rangle$ for the
reasons discussed in Ref.~\cite{KBCGH03}. The constant $A$ is determined by
the condition $AP(t=0,\theta=0 )/ \langle \sigma(E,\theta=0) \rangle =1$.
Figure~\ref{fig1}(a) shows the two slightly overlapping wave packets rotating
towards each other in the backward direction. In panels (b) and (c), the wave
packets strongly overlap around $\theta=180^\circ$, producing interference
fringes. Finally, in panel (d) the wave packets have passed each other and
move apart rotating in the forward direction.

\begin{figure}
\includegraphics[angle=270,width=8.5cm]{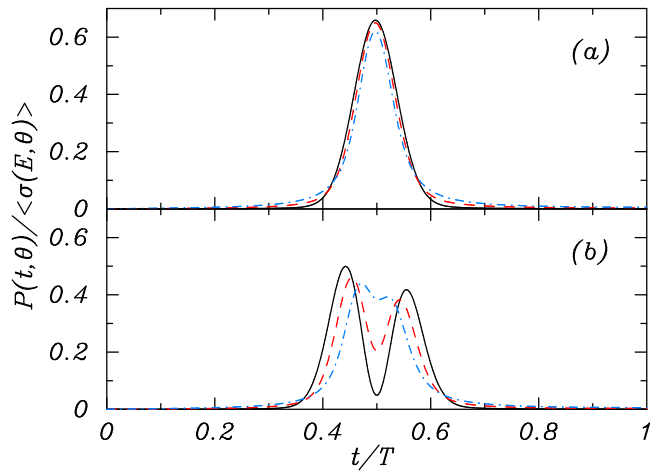}
\caption{\label{fig2} (Color online) Time dependence of the decay
  intensity of the highly excited intermediate complex at
  (a)~$\theta=180^\circ$ and (b)~$\theta=170.6^\circ$, for the
  different sets of $\beta$ and $d$. Solid lines (black) correspond to
  $\beta=0.01$~MeV and $d=3$; dashed (red) to $\beta=0.04$~MeV and
  $d=4$; dashed--dotted (blue) to $\beta=0.075$~MeV and $d=8$. The
  smaller $\beta$ the deeper the minimum is, due to the interference
  between the wave packets for $\theta=170.6^\circ$ at $t=T/2$. }
\end{figure}

In Figs.~\ref{fig1} we also plot $AP(t,\theta )/ \langle
\sigma(E,\theta) \rangle$ with the same constant $A$ and the set of
parameters as before, except for $d$ and $\beta$. We consider two
other cases: $d=4$, $\beta=0.04$~MeV, and $d=8$,
$\beta=0.075$~MeV. One can see that when the wave packets overlap only
slightly (panels (a) and (d)) one can hardly distinguish between the
three combinations of $d$ and $\beta$, even though $\beta$ changes by
almost one order of magnitude. On the other hand, from panels (b) and
(c), we observe that a fringe contrast due to the interference of the
wave packets is very sensitive to absolute values of $\beta$ and
$d$. This result is further illustrated in Fig. 2 where we plot
$P(t,\theta)$ calculated for the three sets of $\beta$ and $d$ values,
as a function of $t$ for $\theta=170.6^\circ$. One observes that the
smaller $\beta$ is the deeper minima are, due to the destructive
interference between the wave packets at $t=T/2$.  However, for
$\theta=180^\circ$, $P(t,\theta)$ in Fig.~\ref{fig2} is insensitive to
the three sets of $\beta$ and $d$.

Note that in order to reproduce the width of the wave packets in
Figs.~\ref{fig1} and~\ref{fig2} for the maximal possible $d=I$ value,
one has to take $\beta=0.1$~MeV. Therefore, even the largest possible
$\beta$ value is more than one order of magnitude smaller than
$\Gamma_{\rm spr}$. The latter is usually estimated from the width of
giant resonances~\cite{BohWei88,GMGW98}: $\Gamma_{\rm spr}\simeq
5$~MeV. This indicates that energy relaxation, i.e., the process of
formation of ergodic many--body configurations, in highly excited
nuclear systems can be much faster than phase relaxation.

\section{Determination of the time power spectrum from the data on the
  excitation function oscillations}

We consider a collision of the spinless reaction partners in the
entrance and exit channels. The cross section of the collision is
given by $\sigma (E, \theta)=|f(E,\theta )|^2$. Here $E$ is the energy
in the entrance channel and $\theta$ is the scattering angle.  The
reaction amplitude $f(E,\theta )$ is a linear combination of
$S$-matrix elements $S^J(E)$ corresponding to the different total spin
values $J$.  Suppose that, for a fixed energy $E_1$, a measurement of
detailed angular distribution is performed. Then, using, {\sl e.g.},
the method~\cite{S-mat96}, one can find $S^J(E_1)$ and, therefore,
$f(E_1,\theta )$. Measuring the angular distributions for different
energies on the energy interval $I=E_{min}-E_{max}$ with the energy
step $\Delta E$ and energy resolution smaller or about the minimal
characteristic energy scale of a variation of $f(E,\theta )$ one
obtains a detailed energy and angle dependencies of the collision
amplitude. Its Fourier component,
\begin{widetext}
  \begin{equation} 
   {\cal P}(t,\theta )\propto
    \int_{E_{min}}^{E_{max}}dE\exp(-iEt/\hbar ) f(E,\theta )\approx
    \Delta E\sum_{n=0}^N \exp[-i(E_{min}+n\Delta E)t/\hbar ]
    f(E_{min}+n\Delta E,\theta ),
\label{eq2}
\end{equation}
\end{widetext}
is the instantaneous decay amplitude received at a detector at time
$t$ with the time uncertainty of about $\hbar/I$, for a collision
initiated at $t=0$.  In the sum in Eq.~(\ref{eq2}), $N=I/\Delta E $ and
$(N+1)$ is the number of equidistant energy values for which the
detailed angular distributions should be measured. The summation in
the r.h.s. of Eq.~(\ref{eq2}) may be changed to the integration if we use a
linear interpolation for $f(E,\theta )$ in between the consequent
energies $E_n$ and $E_{n+1}$. This is justified provided the energy
step $\Delta E$ is smaller or about the minimal characteristic energy
scale of a variation of the cross section. For example, in
Ref.~\cite{Mermaz81} the energy step is $\Delta E_{cm}=133$~keV while
the minimal characteristic scale of the variation of $\sigma
(E,\theta=180^\circ )$ is $\Gamma=0.3$~MeV~\cite{KVG01,KVCG02}.

Having reconstructed ${\cal P}(t,\theta )$ from the excitation
function data one obtains the time power spectra $P(t,\theta )=|{\cal
  P}(t,\theta )|^2$ for any angle. The later quantity is equivalent to
that directly measured using real--time methods of femtochemistry,
with the energy bands of $I$ for the pump and probe laser pulses, to
monitor chemical reactions dynamics~\cite{ZewFemt1-2,Zew03}.  Due to
its dependence on $\theta$, $P(t,\theta )$ determined by the proposed
method could provide even more detailed information on the collision
dynamics, as compared with the femtochemistry experiments, since
latter measurements often allow to obtain $P(t,\theta )$ averaged over
$\theta$.

For the energy interval $I$ being long enough to provide a sufficient
time resolution $\hbar /I$ to resolve interference
fringes~\cite{BKWQDPLB05} the proposed experiment would allow to test
our theoretical predictions. If these were confirmed by the data, then
the experimentally determined $P(t,\theta )$ can be compared with our
calculations presented in Figs.~\ref{fig1} and~\ref{fig2} with the
purpose of evaluation $\beta$ and $d$.  In particular, the energy
interval of 10.5~MeV, over which excitation function for the
$^{12}$C+$^{24}$Mg elastic scattering at $\theta=180^\circ$ was
measured~\cite{Mermaz81}, should be sufficient to resolve the
interference fringes in $P(t,\theta)$~\cite{BKWQDPLB05}.

It should be noted that the method proposed above allows to
reconstruct $P(t,\theta )$ which includes both potential scattering,
{\sl i.e.} fast processes, and the time delayed mechanism. Since the
previous analysis~\cite{KVG01,KVCG02} demonstrated a presence of a
strong potential scattering component we expect that this method will
yield a strong potential scattering contribution in
$P(t,\theta)$. However this contribution is expected to be restricted
to relatively short times $t_{ps}\leq \hbar/\Delta E_{ps}$, where
$\Delta E_{ps}$ is a characteristic energy interval for the variation
of the potential scattering amplitude. For $\Delta E_{ps}\simeq
2-3$~MeV~\cite{GPS95}, $t_{ps}\simeq 3-2\times 10^{-22}$~sec.
Therefore, this potential scattering contribution is expected to be
restricted to $t/T \leq 0.1$ in Fig.~\ref{fig2} and should not show up
at all in Fig.~\ref{fig1}.

The above method to obtain $P(t,\theta )$ from the data is model
independent and, therefore, most reliable one. However it is very
demanding experimentally since $\sigma (E, \theta)$ must be measured
for a very large number of angles~\cite{S-mat96}.  There is another
way to extract $P(t,\theta )$ from the data even if the excitation
function is measured for a single angle. This method can be applied if
the direct reaction process provides a major contribution to the cross
section~\cite{KVCG02}. We consider again a case of spinless collision
partners in the entrance and exit channels. The main idea is based on
the decomposition of the collision amplitude,
$f(E,\theta)=f_{dir}(E,\theta) +\delta f(E,\theta)$. Here
$f_{dir}(E,\theta)$ is the energy smooth amplitude corresponding to
potential scattering or direct fast process, and $\delta f(E,\theta)$
is the energy fluctuating amplitude, $\langle\delta f(E,\theta)\rangle
= 0$, corresponding to the time--delayed collision. Consider first an
idealized case of energy independent amplitude $f_{dir}(E,\theta)$
when it coincides with the energy averaged amplitude
$\langle f(E,\theta)\rangle$.  The collision cross section has the form
$\sigma (E,\theta )=|f(E,\theta)|^2=\sigma_{dir}(E,\theta )+
\sigma_{fl}(E,\theta )+ 2 {\rm Re}[\delta
f(E,\theta)f_{dir}(E,\theta)^\ast ]$. Here the potential scattering or
direct reaction cross section $\sigma_{dir}(\theta
)=|f_{dir}(E,\theta)|^2$ is energy--independent and
$\sigma_{fl}(E,\theta )= |\delta f(E,\theta)|^2$. For a relative
contribution of the potential scattering or direct reaction cross
section to be about 70$\%$ or more of the total cross section,
$\Delta\sigma (E,\theta )=\sigma (E,\theta )-\langle\sigma (E,\theta
)\rangle\simeq 2 {\rm Re}[\delta f(E,\theta)f_{dir}(E,\theta)^\ast ]$,
where $\langle\sigma (E,\theta )\rangle = \sigma_{dir}(\theta )+
\langle\sigma_{fl}(E,\theta )\rangle$ is the energy averaged cross
section. This means that rapid energy variations of the cross section
originate mostly from interference between the energy fluctuating
$\delta f(E,\theta)$ and the energy independent amplitude
$f_{dir}(E,\theta)$. Suppose that the excitation function is measured
on energy interval $I=E_{max}-E_{min}$ with energy step $\Delta E$ so
that the total number of steps is $N+1$, where $N=I/\Delta E$. Then,
for $t > 0$, we have~\cite{KVCG02}
\begin{widetext}
\begin{eqnarray}
  \int_{E_{min}}^{E_{max}}dE\exp(-iEt/\hbar )\Delta\sigma (E,\theta )
  &\propto& \int_{E_{min}}^{E_{max}}dE\exp(-iEt/\hbar )\delta
  f(E,\theta )\nonumber\\
  & & \approx \Delta E\sum_{n=0}^N \exp[-i(E_{min}+n\Delta E)t/\hbar ]
  \delta f(E_{min}+n\Delta E,\theta ) \propto {\cal P}(t,\theta )
\label{eq3}
\end{eqnarray}
\end{widetext}
with $P(t,\theta )=|{\cal P}(t,\theta )|^2$. In Eq.~(\ref{eq3}) we
have employed a causality condition, $ {\cal P}(t < 0,\theta )=0$.
This condition implies that the molecule cannot decay before it is
formed at $t=0$. Again the above information on $P(t,\theta )$ is
equivalent to that obtained directly using real--time methods of
femtochemistry to monitor unimolecular chemical reactions, with the
energy bands of $I=(E_{max}-E_{min})$ for the pump and probe laser
pulses~\cite{ZewFemt1-2}.

The above consideration can be extended to a case when the potential
scattering or direct reaction amplitude $f_{dir}(E,\theta )$ depends
on energy but this dependence is considerably smoother than the energy
dependence of the amplitude $\delta f(E,\theta )$ for the
time--delayed collision. We shall again assume that a relative
contribution of direct processes into the energy averaged cross
section is about 70$\%$ or more. We define the characteristic energy
interval $I_d$ of a variation of $f_{dir}(E,\theta )$ with $I_d\gg
\Gamma $, where $\Gamma$ is the characteristic energy interval of
variation of $\delta f(E,\theta )$. The energy interval $I_d$ can be
evaluated using, {\sl e.g.}, the trend reduction
method~\cite{Papp64}. In particular, for the $^{12}$C+$^{24}$Mg
scattering~\cite{Mermaz81} analyzed in Sec.~II, this interval was
evaluated to be about $4-5$~MeV~\cite{GhoshS87}, {\sl i.e.} much
greater than $\Gamma=0.3$~MeV.  Then one can find the best polynomial
fit of the energy dependence of the cross section $\sigma (E,\theta
)=\sigma_{d}(E,\theta )+ \sigma_{fl}(E,\theta )+ 2 {\rm Re}[\delta
f(E,\theta)f_{dir}(E,\theta)^\ast ]$ with the order of polynomial
being $[I/I_d]+1$, where $[I/I_d]$ is integer part of $I/I_d$ The
resulting energy smooth cross section is denoted as $\tilde \sigma(E,
\theta )\simeq \sigma_d(E, \theta ) + \langle\sigma_{fl}(E, \theta
)\rangle$ with $\langle\sigma_{fl}(E, \theta) \rangle$ being energy
averaged, {\sl i.e.} energy independent, cross section of the
time--delayed processes. It can be found using the standard procedure
employed for the analysis of Ericson fluctuations (see, {\sl e.g.},
Refs.~\cite{GhoshS87,Mermaz81}). As a result, one finds energy smooth
direct reaction cross section $\sigma_{d}(E,\theta )$.  Consider
$\sigma (E, \theta )- \tilde \sigma(E, \theta )\simeq 2 {\rm
  Re}[\delta f(E,\theta)f_{dir}(E,\theta)^\ast ]$.  Assume first that
energy dependence of the phase of $f_{dir}(E,\theta)$ is negligible,
$arg[f_{dir}(E,\theta)]={\rm constant}$, and the smooth energy
dependence of $\sigma_{d}(E,\theta )$ is due to the energy dependence
of $|f_{dir}(E,\theta)|= \sigma_{d}(E,\theta )^{1/2}$. Then applying
the same arguments as those used to obtain Eq.~(\ref{eq3}), for $t>0$, we have
\begin{widetext}
  \begin{equation} {\cal P}(t,\theta )\propto \sum_{n=0}^N
    \exp[-i(E_{min}+n\Delta E)t/\hbar ] [\sigma (E_{min}+n\Delta E,
    \theta )- \tilde \sigma(E_{min}+n\Delta E, \theta
    )]/\sigma_{d}(E_{min}+n\Delta E,\theta )^{1/2}
\label{eq4}
\end{equation}
\end{widetext}
with $P(t,\theta )=|{\cal P}(t,\theta )|^2$.

In order to generalize Eq.~(\ref{eq4}) by taking into account energy
dependence of $arg[f_{dir}(E,\theta)]\equiv \phi(E, \theta)$
($f_{dir}(E,\theta)=|f_{dir}(E,\theta)| \exp[i\phi(E, \theta)]$) we
use the linear approximation $\phi(E, \theta)=\phi(\bar E,
\theta)+(E-\bar E )d\phi(E, \theta)/dE|_{E=\bar E }$, where $\bar
E=(E_{max}-E_{min})/2$ and $d\phi(E, \theta)/dE|_{E=\bar E
}=t_{dir}/\hbar$ with $t_{dir}\ll \hbar/\Gamma$ being the time delay
due to the potential scattering or direct reactions~\cite{Wigner55}.
Then, it is easy to see, that, for $t>t_{dir}$, the r.h.s. of
Eq.~(\ref{eq4}) is changed to ${\cal P}(t-t_{dir},\theta )$.  This
manifestly demonstrates that the interference between time--delayed
and fast direct processes in the cross section is a precondition for
monitoring the time evolution, the fast process switches on the clock
at the initial moment of time $t_d$ playing the role of the pump
pulse. In the absence of direct processes the initial moment of time
is not defined.

\begin{figure}
\includegraphics[angle=270,width=8.5cm]{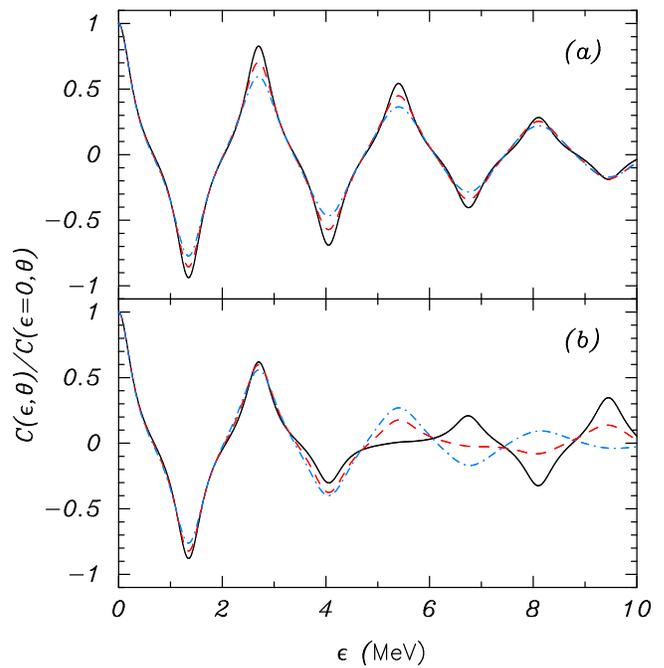}
\caption{\label{fig3} (Color online) Directly measurable cross section
  energy autocorrelation functions calculated at
  (a)~$\theta=180^\circ$ and (b)~$\theta=170.6^\circ$ for the
  different sets of $\beta$ and $d$. Solid lines (black) corresponds
  to $\beta=0.01$~MeV and $d=3$; dashed (red) to $\beta=0.04$~MeV and
  $d=4$; dashed-dotted (blue) to $\beta=0.075$~MeV and $d=8$.  }
\end{figure}

Note that $P(t,\theta )$ can also be expressed as a cosine
half--Fourier transform of the cross--section energy autocorrelation
function $C(\varepsilon,\theta )$, provided the relative contribution
of potential scattering is larger than $70\%$~\cite{KVG01,KVCG02}. In
Fig.~\ref{fig3} we plot $C(\varepsilon,\theta )/C(\varepsilon=0,\theta
)= [\int_0^\infty dt \cos (\varepsilon t/\hbar) P(t,\theta ) ]/
\langle \sigma(E,\theta) \rangle $ for (a)~$\theta=180^\circ$ and
(b)~$\theta=170.6^\circ $, for the three combinations of $\beta$ and
$d$. Since the $P(t,\theta=180^\circ )$ are close for the different
sets of $\beta$ and $d$, the corresponding
$C(\varepsilon,\theta=180^\circ )$ can hardly be distinguished
reliably. This is the reason that the analysis at
$\theta=180^\circ$~\cite{KVG01,KVCG02} is not sufficient to determine
unambiguously the values of $\beta$ and $d$. However,
$C(\varepsilon,\theta=170.6^\circ )$ is more sensitive to the
different sets of $\beta$ and $d$. In particular, for $\varepsilon\geq
4-5$~MeV, the oscillations in $C(\varepsilon,\theta=170.6^\circ )$ for
$\beta=0.01$~MeV, $d=3$ and $\beta=0.075$~MeV, $d=8$ are out of phase
with the absolute value of their difference being up to $0.4$.  Still,
from the comparison of Figs.~\ref{fig1}(c), \ref{fig2}(b) and
Fig.~\ref{fig3} we observe that the sensitivity of $P(t,\theta)$ to
different values of $\beta$ and $d$ is considerably stronger than the
sensitivity of $C(\varepsilon,\theta )$ to these values. Also
$P(t,\theta)$ obtained using Eqs.~(\ref{eq3}) and~(\ref{eq4}) does not
acquire statistical errors due to the finite energy interval
$I$. Indeed such a method provides information on $P(t,\theta)$
equivalent to that obtained in the femtochemistry
experiments. Therefore the only uncertainty of this method is a finite
time resolution, $\Delta t \simeq \hbar/I$, which is the uncertainty
in the femtochemistry experiments with the finite energy band $I$ of
the pump and probe laser pulses~\cite{Zew03,ZewFemt1-2}.  On the other
hand, a reconstruction of $P(t,\theta )$ from $C(\varepsilon,\theta )$
may result in additional errors due to possible statistical
uncertainties in $C(\varepsilon,\theta )$ related to the finite energy
range $I$~\cite{Richt74}. Such additional statistical uncertainties
may not be excluded even though the effects studied in this paper can
clearly not be associated with Ericson fluctuations. Indeed, unlike,
{\sl e.g.}, oscillating structures in Fig.~\ref{fig3}, Ericson theory
predicts angle independent Lorentzian shapes for the
$C(\varepsilon,\theta )$. Therefore, for the reasons described above,
we suggest that a reconstruction of $P(t,\theta)$ directly from the
excitation functions (Eqs.~(\ref{eq3}) and~(\ref{eq4})) is more
reliable than from $C(\varepsilon,\theta )$.

\section{From heavy--ion collisions to bimolecular chemical 
reactions}

An important question is whether the rotating wave packets and their
interference can occur in colliding systems for which more rigorous
approaches can be implemented. The encouraging answer has been given
by the calculations of the time evolution of the $H+D_2\to HD+D$
bimolecular chemical reaction~\cite{Nat02,Alt1,Alt2}. The evidence for
rotational wave packets has been demonstrated (see, e.g., Fig.~3
in~\cite{Nat02}) due to the interference of the overlapping
resonances~\cite{Chao01} of the quasibound complex with $J=15-20$
total spin values, (i.e., within our approach, $d\simeq 5$). In this
peripheral time--delayed collision the rotational wave packets are
stable showing a moderate spreading during half a period of the
rotation. Within our description, this reveals a small $\beta$ value
in the time--delayed peripheral reaction mechanism, {\sl i.e.}
$\pi\beta/\hbar\omega \ll 1/d$. Then, in accordance with our
prediction, the coherent superposition of the rotating wave packets
should produce a strong interference contrast. This is indeed the case
as clearly seen in Fig.~3 of Ref.~\cite{Nat02}. Similar effects of the
stable rotating wave packets and their coherent superpositions with
strong interference contrast have also been found in the calculations
for $F+HD\to HF+D$ bimolecular reaction~\cite{Alt3,Alt4}.

The above results should allow to test our suggestion for the
determination of the phase relaxation time and, in general, for the
time evolution of the collision in bimolecular chemical reactions. For
example, in~\cite{Alt1} a total time dependent amplitude was
calculated for the $H+D_2\to HD+D$ reaction. Then it was splitted into
short--time direct reaction component and the time--delayed
one. Taking the half--Fourier transform, explicit energy and angle
dependencies of both the direct and time--delayed amplitudes were
obtained. It was found that in the region of backward angles
$\theta\leq 70^\circ$ and for the c.m. energy interval 1.6--2~eV the
direct reaction cross section is considerably bigger than the
time-delayed one. Then one can apply our method (Sect. III) to obtain
$P(t,\theta )$ of the time--delayed reaction mechanism from the energy
dependence of the double--differential cross sections. This extracted
$P(t,\theta )$ can be compared with its actual exact form since the
latter is known and was used to generate the energy dependence of the
double--differential cross sections.

The calculations~\cite{Alt1,Alt2} suggests that in the forward angle
range, where the Schr\"odinger cat states originated from the
interference of the rotating towards each other wave packets are
observed, contribution of the direct reactions is
negligible. Therefore, for this particular $H+D_2\to HD+D$ reaction
for the c.m. energy less than 2~eV our method can not be applied.  Yet
one can still check the sensitivity of the method by artificially
adding an energy smooth background amplitude to the calculated one for
the time--delayed reaction mechanism. The energy smooth amplitude
would play a role of the direct reaction amplitude. This would enable
one to directly test the accuracy of the determination of the
interference fringe contrast from the data on the double differential
cross sections generated with the artificially added energy smooth
direct reaction amplitude. Such a test may be important since for the
higher energy and/or for some other bimolecular chemical reactions, a
strong contribution of direct reactions is possible for the same angle
range where the Schr\"odinger cat states show up.

Note that pure energy resolution measurements of the cross sections of
bimolecular chemical reactions, in particular $H+D_2\to
D+HD$~\cite{Nat03} and $F+HD\to HF+D$~\cite{Skod} reactions, have
recently become possible. Our results suggest that this should enable
experimentalists to extract information on the time dependence, in
particular on the rotational coherent dynamics and possible wave
packet interference, of the bimolecular chemical reactions which
previously could only be studied by the pump probe laser pulses
technique to monitor unimolecular chemical
reactions~\cite{ZewFemt1-2,Zew03}.

\section{Phase relaxation versus decoherence}

Since phase relaxation results in washing out the interference fringes
and eventually, for a sufficiently large $\beta$, can destroy the
Schr\"{o}dinger cat states of Figs.~\ref{fig1} and~\ref{fig2}, this
effect appears to be quite similar to decoherence~\cite{Myatt00}. Yet,
there is an essential difference between the two effects. It is
instructive to compare the extreme case of very fast phase relaxation
(very large $\beta$) with very fast decoherence. Very fast phase
relaxation corresponds to the regime of random matrix theory, {\sl
  i.e.}, to random phase relations between all partial width
amplitudes corresponding to different strongly overlapping resonance
states~\cite{GMGW98}. This yields angle independent exponential form
for the time power spectrum $P(t,\theta )$ and unity for the
normalized variance of the fluctuating cross section. Analogous to the
fast phase relaxation would be quick decoherence between all
many--body quasi--stationary resonance states, similar to the
decoherence of Fock states due to the coupling of the system to a
phase reservoir~\cite{Myatt00}. Such a decoherence would destroy
interference terms between all the many--body eigenstates, leading to
the vanishing of interference between different partial width
amplitudes. It can be shown that in this case the time power spectrum
would have the same angle independent exponential form as for fast
phase relaxation.  However, in contrast to fast relaxation, quick
decoherence would result in washing out the cross section energy
fluctuations, reducing its normalized variance to $D/\Gamma\ll
1$~\cite{Kun02}. It should be noted that the suppression of electron
transmission intensity fluctuations through nanostructures due to
dephasing, an effect similar to decoherence, can be described on the
basis of different arguments allowing to evaluate dephasing rates from
the amplitude of conductance fluctuations~\cite{Marcus93}.

\section{Discussion and conclusions}

Our approach is based on a new idea of slow phase relaxation between
partial width amplitudes in a regime of strongly overlapping
resonances. Contrary to the conventional idea of random phase
relations in this regime~\cite{GMGW98,Zewail83,Smil}, our analysis
invokes spin off--diagonal phase correlations and their slow decay for
the resonance time--delayed collisions. Though these effects may occur
for atomic cluster collisions~\cite{KBCGH03} and are already revealed
for the bimolecular chemical
reactions~\cite{Nat02,Alt1,Alt2,Alt3,Alt4}, the present consideration
is essentially motivated by the pure energy resolution data on
fluctuations in $^{12}$C+$^{24}$Mg elastic {\it and} inelastic
scattering at $\theta=180^\circ$~\cite{GhoshS87,Mermaz81},
demonstrating both the fine energy structure of $\simeq 0.3$~MeV and
quasiperiodicity with period $\simeq 3$~MeV~\cite{KVG01,KVCG02}. It is
these peculiar features that led us to the interpretation of the
$^{12}$C+$^{24}$Mg scattering in terms of highly excited
quasimolecular states with strongly overlapping resonances, stable
rotational wave packets and strong sensitivity of their coherent
superpositions to the phase relaxation width. Note that quasiperiodic
structures in the excitation functions are present also for other
heavy--ion colliding systems~\cite{KRV99}. We have also described
thoroughly how to extract the time power spectrum from the data on the
excitation function oscillations.

It should be mentioned that a conceptually different optical model
description of the $^{12}$C+$^{24}$Mg elastic scattering at backward
angles was presented in Ref.~\cite{Lep99}, where the quasimolecular
structures were ruled out. This conclusion was based on a qualitative
description of angular oscillations at backward angles for only three
energies within the same energy range which is the subject of our
study in this paper. Unfortunately, that analysis is inconsistent with
the presence of fine energy structure of $\simeq 0.3$~MeV at
$\theta=180^\circ$~\cite{GhoshS87}, which unambiguously reveals the
presence of time--delayed processes at backward angles. On the
contrary, the optical model description relies on the energy average
$S$--matrix and therefore is expected to reproduce broad $\simeq
2-3$~MeV energy structures in the cross section. From our point of
view, the fact that the presence of the fine energy structures was not
taken into account led to an incorrect qualitative and then
quantitative identification of the physical picture of the collision
process in Ref.~\cite{Lep99}.

In conclusion, we have demonstrated a strong sensitivity of
Schr\"odinger cat states to the phase--relaxation width in complex
quantum collisions. This should permit to determine the phase
relaxation time from measurements of the excitation functions for
$^{12}$C+$^{24}$Mg elastic scattering at backward angles. Such an
experiment would be desirable since the Schr\"odinger cat states
predicted in~\cite{KBCGH03} for $^{12}$C+$^{24}$Mg scattering involve
$\sim 10^3 -10^4$ many--body configurations of the highly excited
intermediate complex. To the best of our knowledge, the internal
interactive complexity of these quantum macroscopic superpositions
dramatically exceeds~\cite{KBCGH03} all those previously
experimentally realized.  The proposed method can also be applied for
determination of the phase relaxation time from the data on excitation
functions for bimolecular chemical reactions.

\begin{acknowledgments}
We are grateful to F. Leyvraz and T.H. Seligman for useful
discussions and suggestions. One of us (LB) acknowledges financial support
from the projects IN--101603 (DGAPA--UNAM) and 43375--E (CONACyT).
\end{acknowledgments}

\end{document}